\documentclass[
    ,final            
  ]
  {aipproc}
\usepackage{epsfig}
\usepackage{axodraw}
\usepackage{graphicx}
\usepackage{amssymb}
\layoutstyle{6x9}

\newcommand{\la}{{\lambda}}

\newcommand{\be}{\begin{equation}}
\newcommand{\ee}{\end{equation}}
\newcommand{\beq}{\begin{equation}}
\newcommand{\eeq}{\end{equation}}
\newcommand{\bea}{\begin{eqnarray}}
\newcommand{\eea}{\end{eqnarray}}
\newcommand{\br}{\begin{eqnarray}}
\newcommand{\er}{\end{eqnarray}}
\newcommand{\ba}{\begin{array}}
\newcommand{\ea}{\end{array}}
\newcommand{\bi}{\begin{itemize}}
\newcommand{\ei}{\end{itemize}}
\newcommand{\bn}{\begin{enumerate}}
\newcommand{\en}{\end{enumerate}}
\newcommand{\bc}{\begin{center}}
\newcommand{\ec}{\end{center}}

\def\bY{{\bf Y}}
\def\bA{{\bf A}}

\def\bU{{\bf U}}

\def\bmm{{\bf m}}

\def\tl{{\tilde{L}}}

\def\te{{\tilde{e^c}}}

\def\tq{{\tilde{Q}}}
\def\tu{{\tilde{u^c}}}

\def\unity{{\hbox{1\kern-.8mm l}}}
\newcommand{\ov}{\overline}

\newcommand{\no}{\nonumber}

\newcommand{\ga}{\gamma}
\newcommand{\gsim}{\lower.7ex\hbox{$\;\stackrel{\textstyle>}{\sim}\;$}}
\begin{document}

\title{Relating seesaw neutrino masses, lepton flavor violation and SUSY breaking}


\author{Filipe R. Joaquim}{
address={Istituto Nazionale di Fisica Nucleare (INFN), Sezione di
Padova, I-35131 Padua, Italy},altaddress={Dipartimento di Fisica
``G.~Galilei'', Universit\`a di Padova I-35131 Padua, Italy}}

\author{Anna Rossi}{
  address={Dipartimento di Fisica ``G.~Galilei'', Universit\`a di
Padova I-35131 Padua, Italy} }


\begin{abstract}
We discuss a GUT realization of the supersymmetric triplet seesaw
mechanism (recently proposed by us in hep-ph/0604083 and further
analyzed in hep-ph/0607298) where the exchange of the heavy triplet
states generates both neutrino masses and soft SUSY breaking terms.
\end{abstract}

\maketitle

Recently, we have proposed a novel supersymmetric scenario of the
triplet seesaw mechanism where the soft SUSY breaking (SSB)
parameters of the minimal supersymmetric standard model (MSSM) are
generated at the decoupling of the heavy triplets~\cite{af1}. The
mass scale of such SSB terms is fixed {\it exclusively} by the
triplet SSB bilinear term $B_T$ and flavor violation (FV) in the SSB
MSSM lagrangian can be directly related with the low-energy neutrino
parameters. Our scenario is therefore highly predictive since it
relates neutrino masses, lepton flavor violation (LFV) in the
sfermion sector and electroweak symmetry breaking (EWSB).

We consider an $SU(5)$ grand unified theory (GUT) where the triplet
states $T\sim (3,1)$ and $\bar{T} \sim (3,-1)$ fit into the 15
representation $15 = S + T +Z$ transforming as $ S\sim
(6,1,-\frac23), ~T \sim (1,3,1),~ Z\sim (3,2,\frac16)$ under $SU(3)
\times SU(2)_W \times U(1)_Y$ (the $\ov{15}$ decomposition is
obvious).
The SUSY breaking mechanism is triggered by a gauge singlet chiral
supermultiplet $X$, whose scalar $S_X$ and auxiliary $F_X$
components are assumed to acquire a vacuum expectation value.
Defining the $B-L$ quantum numbers of the various fields as being a
combination of their hypercharges and the charges:
\be
Q_{10}= \frac15\;,\; Q_{\bar{5}} = - \frac35\;,\;Q_{{5}_H}
=-\frac25\;,\;Q_{\bar{5}_H }= \frac25\;,\; Q_{15} =
\frac65\;,\;Q_{\ov{15}}=
\frac45,\,Q_{X} = -2\,,\label{charges}\ee %
and imposing $B-L$ conservation, the $SU(5)$ superpotential reads
\bea W_{SU(5)}&=& \frac{1}{\sqrt2}(\bY_{15} \bar5~ 15 ~\bar5 +
 \la {5}_H ~\ov{15}~ {5}_H)
 + \bY_5 \bar5 ~ \bar 5_H 10 + \bY_{10} 10 ~10 ~5_H  \no \\
&&+ M_5  5_H~\bar5_H + \xi X 15 ~ \ov{15} \,, \label{su5} \eea
where we have used the usual conventions for the $SU(5)$
representations. It is clear from $W_{SU(5)}$ that the $15, \ov{15}$
states act as {\it messengers} of both $B-L$ and SUSY breaking to
the visible (MSSM) sector due to the coupling with $X$. In
particular, while $\langle S_X\rangle$ only breaks $B-L$, $\langle
F_X\rangle$ breaks both SUSY and $B-L$.
Once $SU(5)$ is broken to the SM group we find, below the GUT scale
$M_G$, $W = W_0 + W_T + W_{S,Z} $ with,
 \bea
&&W_0 =  \bY_e  e^c H_1  L +\bY_d d^c H_1  Q + \bY_u  u^c Q  H_2  +
\mu H_2 H_1\,,\no \\
&& W_T= \frac{1}{\sqrt{2}}(\bY_{T} L T L  + \la H_2 \bar{T} H_2) +
  M_T T \bar{T}\,,\label{WT} \no \\
&&W_{S,Z}= \frac{1}{\sqrt{2}}\bY_S d^c S d^c + \bY_Z  d^c  Z L + M_Z
Z\bar{Z}+M_S S\bar{S}
.\label{su5b} \eea%
Here, $W_0$ denotes the MSSM superpotential, the term $W_T$ contains
the triplet Yukawa and mass terms, and $W_{S,Z}$ includes the
couplings and masses of the colored fragments $S,Z$. For simplicity,
we take $M_T = M_S = M_Z$ and $\bY_{S},\bY_{Z}\ll \bY_T$ at $M_{G}$
(the general $SU(5)$ has been studied in detail in Ref.~\cite{af2}).
In Eq.~(\ref{su5b}), $W_T$ is responsible for the realization of the
seesaw mechanism. The Majorana neutrino mass matrix reads, at the
electroweak scale,
 \be \label{T-mass}
{\bf m}^{ij}_\nu =\frac{ \la\langle H_2\rangle^2}{M_T} \bY^{i j}_T ,
~~~ i,j=e,\mu,\tau . \ee
In the basis where $\bY_e$ is diagonal, it is apparent that all LFV
is encoded in $\bY_T$. This stems from the fact that the nine
independent parameters contained in $\bmm_\nu$  are directly linked
to the neutrino parameters according to $\bmm_\nu = \bU^* \bmm^D_\nu
\bU^\dagger$, where $\bmm^D_\nu = {\rm diag}(m_1,m_2, m_3)$ are the
mass eigenvalues, and $\bU$ is the leptonic mixing matrix.

As for the SSB term one has, in the broken phase, $-B_T M_T
(T\bar{T} + S\bar{S} + Z\bar{Z}) + {\rm h.c.}$, with $B_T \equiv
B_{15}$. These terms lift the tree-level mass degeneracy in the MSSM
supermultiplets. Indeed, at the scale $M_T$, all the states $T,
\bar{T}, S, \bar{S}$ and $Z, \bar{Z}$ are {\it messengers} of SUSY
breaking to the MSSM sector via gauge interactions, as it happens in
conventional gauge-mediation scenarios. However, in our scenario the
states $T, \bar{T}$ also communicate SUSY-breaking through Yukawa
interactions. Finite contributions  for the trilinear couplings of
the superpartners with the Higgs doublets, $\bA_e, \bA_u, \bA_d$,
the gaugino masses $M_a~ (a=1,2,3)$ and the Higgs bilinear term $-
B_H\mu H_2 H_1$ emerge at the one-loop level:%
\be
 {\bA}_e  =
\frac{3 B_T}{16 \pi^2} \bY_e \bY^\dagger_T \bY_T\;\;,\;\;
 {\bA}_u =
\frac{3 B_T|\la|^2}{16 \pi^2} \bY_u  \;\;,\;\;
 {\bA}_d = 0\;\;,\;\;
 M_a =   \frac{7 B_T g_a^2}{16 \pi^2}\;\;,\;\;
B_H = \frac{3 B_T|\la|^2}{16 \pi^2}\,.
 \label{finite}
\ee Instead, the finite
contributions to the scalar SSB masses arise at the two-loop level:
\bea \bmm^2_{\tl} &=&\frac{B_T^2}{(16 \pi^2)^2} \left[
\frac{21}{10} g^4_1 + \frac{21}{2} g^4_2 - (\frac{27}{5} g^2_1
+21 g^2_2)\bY^\dagger_T\bY_T + 3\bY^\dagger_T \bY^T_e \bY^*_e \bY_T+
18  (\bY^\dagger_T\bY_T)^2\right. \no \\
&& \!\!\!\!\!\!\!\! \left.   + 3{\rm Tr}(\bY^\dagger_T\bY_T)
\bY^\dagger_T\bY_T \right]\quad , \quad \bmm^2_{\te}
=\frac{B_T^2}{(16 \pi^2)^2} \left[\frac{42}{5} g^4_1 - 6 \bY_e
\bY^\dagger_T\bY_T\bY^\dagger_e
\right]\,,\no \\
m^2_{H_2}&=&\frac{B_T^2}{(16 \pi^2)^2} \left[ \frac{21}{10} g^4_1 +
\frac{21}{2}g^4_2
 - \left(\frac{27}{5} g^2_1
+21 g^2_2\right)|\la|^2  + 9 |\la|^2 {\rm Tr}(\bY_u\bY^\dagger_u) +
21 |\la|^4 \right]\,,\no \\
m^2_{H_1} &=& \frac{B_T^2}{(16 \pi^2)^2} \left[
 \frac{21}{10} g^4_1 +   \frac{21}{2}  g^4_2\right]\,.
 \label{soft2} \eea 
In the above equations we have only shown the result for the slepton
and Higgs soft masses $\bmm^2_{\tl}$ and $m^2_{H_{1,2}}$,
respectively. Since they are not directly relevant for our present
discussion, we do not provide here the results for the squark soft
masses $\bmm^2_{\tu}$ and $\bmm^2_{\tq}$ which can be found in
Ref.~\cite{af1}. The expressions in Eqs.~(\ref{finite}) and
(\ref{soft2}) hold at the decoupling scale $M_T$ and therefore are
meant as boundary conditions for the SSB parameters which then
undergo (MSSM) RG running  to the low-energy scale $\mu_{SUSY}$.
In particular, we observe that the  Yukawa couplings $\bY_T$ induce
LFV to $\bA_e$, to the scalar masses $\bmm^2_\tl$ and  $\bmm^2_\te$.
This makes the present scenario distinct from pure gauge-mediated
models where FV comes out naturally suppressed.

The crucial point in our discussion is that the flavor structure of
$\bmm^2_\tl$ is proportional to $\bY^\dagger_T \bY_T$ which can be
written by using Eq. (\ref{T-mass}) in terms of the neutrino
parameters (the terms $\propto g^2 \bY^\dagger_T\bY_T$ are
generically the leading ones): 
\be (\bmm^2_\tl)_{ij} \propto {B_T}^2 (\bY^\dagger_T \bY_T)_{ij} \sim
B_T^2 \left(\frac{M_T}{\la \langle H_2\rangle^2}\right)^{\!\!2}\!\!
\left[\bU (\bmm^{D }_\nu)^2 \bU^\dagger\right]_{ij} . \ee
Consequently, the relative size of LFV in the different leptonic
families  can be univocally predicted as: 
\be \label{predi}
\frac{ (\bmm^{2 }_{\tilde{L}})_{\tau \mu}}
  {(\bmm^{2 }_{\tilde{L}})_{\mu e} } \approx
\left(\frac{m_3}{m_2}\right)^2 \frac{\sin 2\theta_{23}}{\sin
2\theta_{12} \cos\theta_{23}} \sim 40\quad , \quad   \frac{ (\bmm^{2
}_{\tilde{L}})_{\tau e}}
  {(\bmm^{2 }_{\tilde{L}})_{\mu e} } \approx \tan\theta_{23} \sim 1 ,
\ee
where $\theta_{12}$ and $\theta_{23}$ are lepton mixing angles. This
aspect renders the present framework much more predictive than the
type I seesaw mechanism. Indeed, \emph{model-independent relations
like the ones shown above cannot be found in the former case without
making assumptions about the high-energy flavor structure}. From
Eqs.~(\ref{predi}) the branching ratios (BR) of LFV processes such
as the decays $\ell_i \to \ell_j \gamma$ can be predicted
\be\label{brs} {\rm BR}(\tau \to \mu \gamma)/ {\rm BR}(\mu
\to e \gamma) \sim 300 \quad,\quad{\rm BR}(\tau \to e \gamma)/{\rm
BR}(\mu \to e \gamma) \sim 10^{-1}\,,
 \ee
where the estimates have been obtained considering a hierarchical
neutrino mass spectrum and the best-fit values for the low-energy
neutrino oscillation parameters. Relations like those of
Eqs.~(\ref{predi}) and (\ref{brs}) are equally obtained if one
assumes universal boundary conditions for the soft masses at a scale
higher than $M_T$\cite{ar}. It is worth stressing that our scenario
constitutes a concrete and simple realization of the so-called
minimal lepton flavor violation hypothesis.  Other LFV processes and
related correlations have been considered in~\cite{af2}. Without
loss of generality we take $B_T$ to be real\footnote{For discussions
on the possible implications of a complex $B_T$ to electric dipole
moments and the generation of the baryon asymmetry of the Universe
see Refs.~\cite{cmrv} and \cite{lepto}, respectively.}.

Following a bottom-up perspective and taking a given ratio $M_T/\la$
and $\tan\beta$, $\bY_T$ is determined at $M_T$ according to the
matching expressed by Eq.~(\ref{T-mass}) using the low-energy
neutrino parameters. Although the $\mu$-parameter is not predicted
by the underlying theory, it is nevertheless determined together
with $\tan\beta$ by correct EWSB conditions. Therefore, we end up
with only three free parameters, $B_T, M_T$ and $\la$.
\begin{figure}
\begin{tabular}{cc}
\hspace*{-0.2cm}\includegraphics[width=7.3cm]{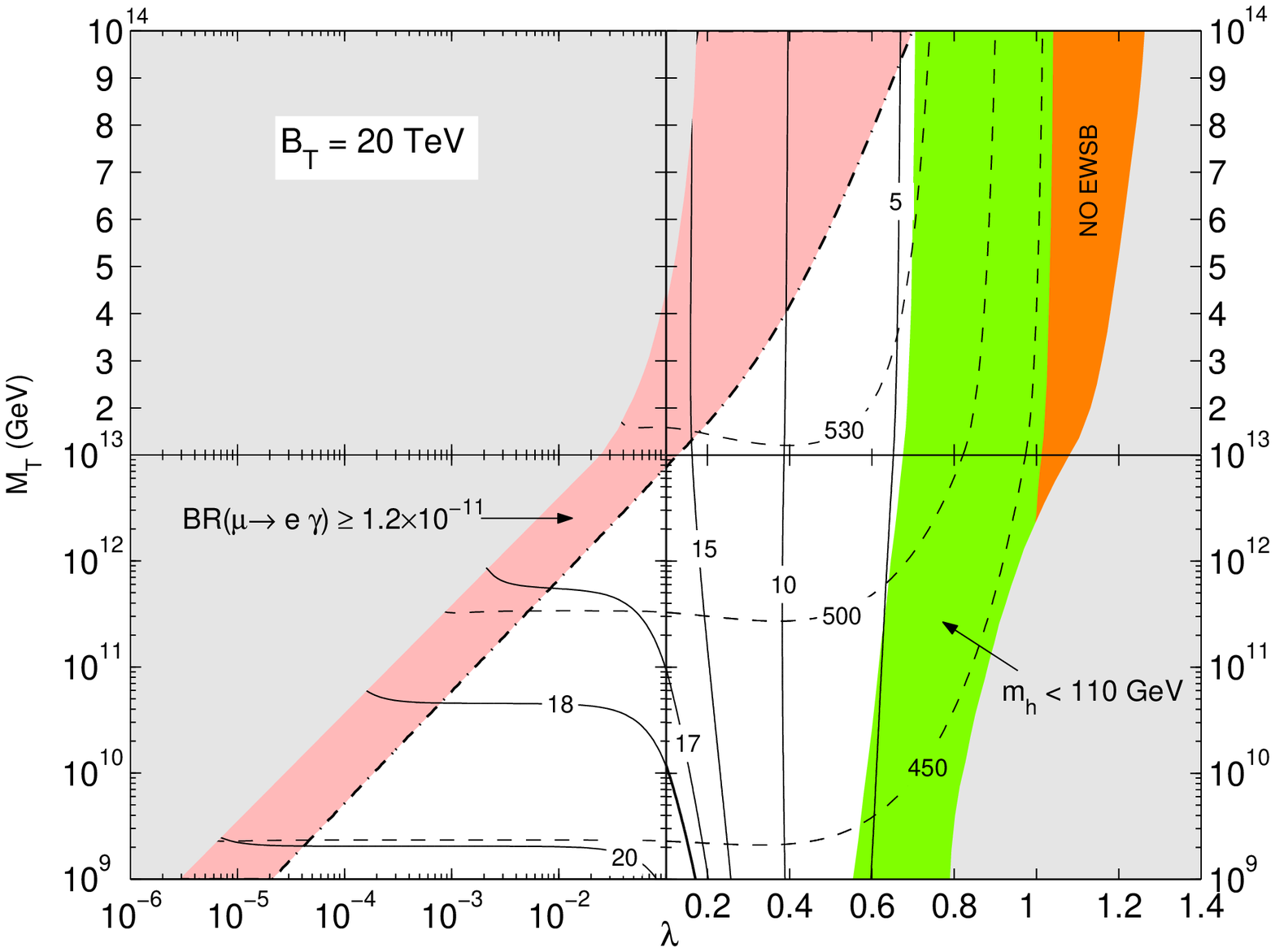}
&\hspace*{-0.5cm}
\includegraphics[width=7.3cm,height=5.4cm]{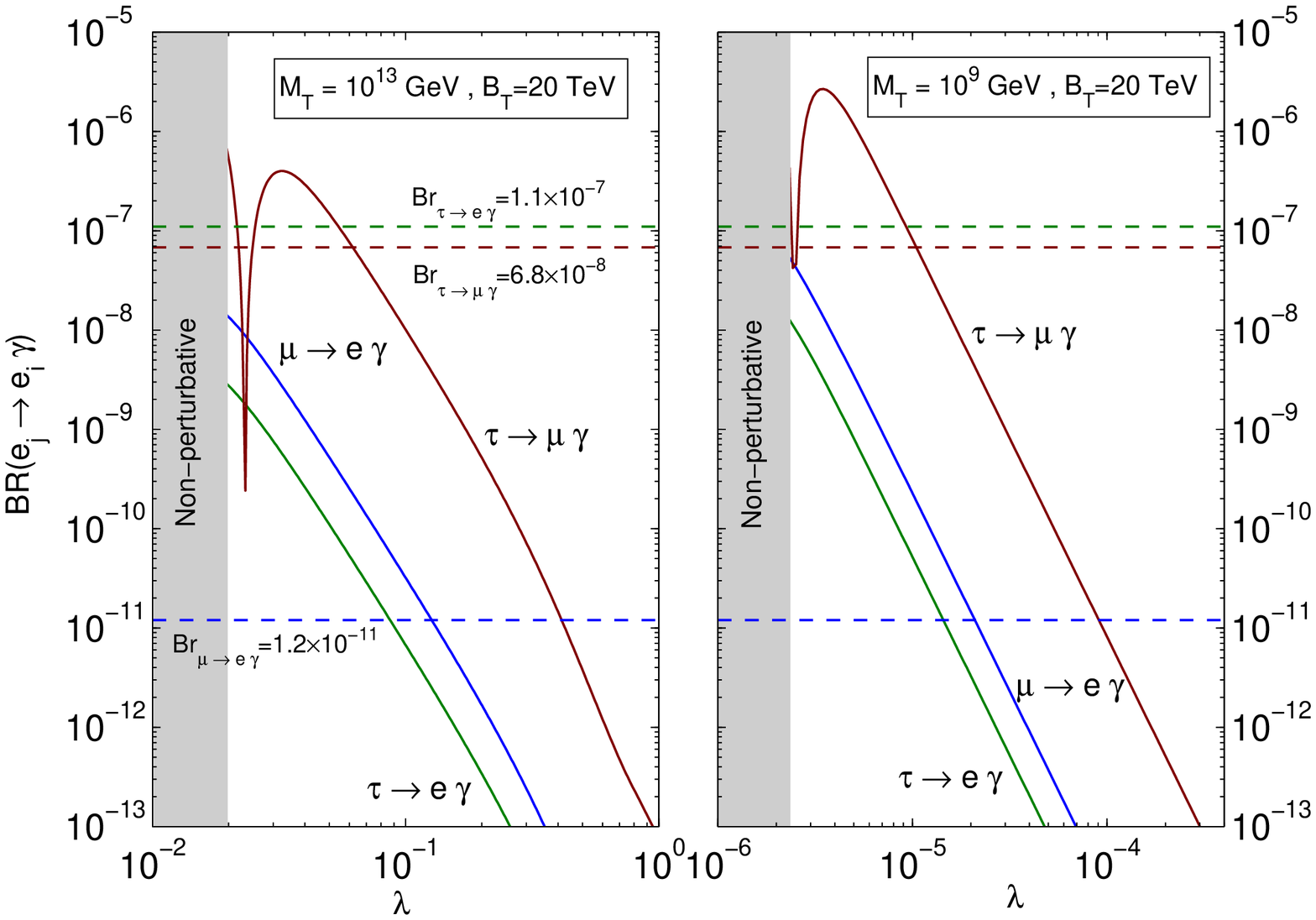}
\end{tabular}
\caption{\it Left-panel: $(\la, M_T)$ parameter space analysis (see
text for more details) for $B_T=20~{\rm TeV}$. The isocontours of
$\tan\beta$ (solid lines) and $\mu$ (dashed lines) are shown.
Right-panel: $BR$s of the lepton radiative decays as a function of
$\la$ for $B_T=20~{\rm TeV}$ and $M_T=10^{13} (10^{9})~{\rm GeV}$ in
the left (right) plot. The horizontal lines indicate the present
bound on each $BR$.} \label{f1}
\end{figure} 

In the left panel of Fig.~\ref{f1} we show the $(\la, M_T)$
parameter space allowed by the perturbativity (lightest grey region)
and EWSB requirements, the experimental lower bound on the lightest
Higgs mass $m_h$ and the upper bound on $BR(\mu \to e \ga)$, for
$B_T = 20~{\rm TeV}$. The white region shows the portion of the
parameter space allowed by the aforementioned constraints (for
extensive discussions on the interpretation of this plot see
Refs.~\cite{af1,af2}). In the right-panel, we display the branching
ratios $BR(\ell_j \to \ell_i \gamma)$ as a function of $\la$ for
$B_T= 20~{\rm TeV}$ and $M_T= 10^{13}\,(10^{9})~{\rm GeV}$ in the
left (right) plot. The behavior of these branching ratios is in
remarkable agreement with the estimates of Eq.~(\ref{brs}). Hence,
the relative size of LFV does not depend on the detail of the model,
such as the values of $\la$, $B_T$ or $M_T$.

In conclusion, we have suggested a unified picture of the
supersymmetric type-II seesaw where the triplets, besides being
responsible for neutrino mass generation, communicate SUSY breaking
to the observable sector through gauge and Yukawa interactions. We
have performed a phenomenological analysis of the allowed parameter
space emphasizing the role of LFV processes in testing our
framework. More details can be found in Refs.~\cite{af1} and
\cite{af2}.\vspace*{-0.1cm}
\begin{theacknowledgments} The
work of F.R.J. is supported by 
FCT-Portugal under the grant \mbox{SFRH/BPD/14473/2003},  INFN and
PRIN Fisica Astroparticellare (MIUR). The work of A.~R.~ is
partially supported by the project EU MRTN-CT-2004-503369.
\end{theacknowledgments}

\end{document}